\documentstyle[12pt]{article}
\begin{document}
\baselineskip 22pt
\begin{center}
{\Large Anonymous key quantum cryptography and unconditionally secure
quantum bit commitment} \\
\vspace*{.2in}

{\Large Horace P. Yuen} \\ 
{\large Department of Electrical and Computer
Engineering \\ Department of Physics and Astronomy\\
Northwestern University \\ Evanston IL  60208-3118 \\ email: yuen@ece.northwestern.edu}
\end{center}
%\widetext
\begin{abstract}
A new cryptographic tool, anonymous quantum key technique, is
introduced that leads to unconditionally secure key distribution and
encryption schemes that can be readily implemented experimentally in a
realistic environment.  If quantum memory is available, the technique
would have many features of public-key cryptography; an identification
protocol that does not require a shared secret key is provided as an
illustration.  The possibility is also indicated for obtaining
unconditionally secure quantum bit commitment protocols with this technique.
\end{abstract}

This paper has the same title as my Capri talk but the contents are
not identical.  The portion on anonymous key is greatly expanded here,
while only brief mention is made on quantum bit commitment, a detailed
treatment of which is available in Ref.~\cite{yuenun}.

A classic goal of cryptography is privacy: two parties wish to
communicate privately so that an adversary can learn nothing about its
content. This was usually achieved through the use of a shared private
key, typically a string of binary digits, for encrypting and
decrypting the message data.  A revolution in cryptography occurred
around 1976 with the emergence of {\em public-key}
cryptography \cite{menezes}, in which knowledge of a public key for
encryption would not lead to knowledge of a secret private key for
decryption.  The concept of digital signature, the binding of a signer
to an electronic digital message, was introduced via public-key
technique.  The idea of using quantum physics for cryptographic
purpose was first proposed by Wiesner in the early
1970's \cite{wiesner}.  It came to fruition in the work of Bennett and
Brassard \cite{bennet1} on key distribution, culminating in an
experimental prototype demonstration \cite{bennet2}.  Despite earlier
papers on the use of quantum cryptography to achieve other
cryptographic goals, it turns out that key distribution is the only
viable one so far \cite{lo}.  Also, the lack of a quantum
authentication scheme implies that some standard classical technique
has to be employed which takes away some of the novelty of the quantum
techniques, which at first sight seem to be public-key type protocols
that do not require the prior sharing of secret information.

Consider two users, Adam and Babe, with a powerful adversary Eve who
can manipulate all the communications between them. In an
intruder-in-the-middle or {\em impersonation} attack, Eve can pretend
to be Adam to Babe, and Babe to Adam, in all the known quantum
protocols.  If Adam and Babe do not have a prior shared secret key for
message authentication, it is often assumed that a non-jammable
classical public channel would prevent impersonation. This, however,
is not the case \cite{bb84} as there is still a user authentication
({\em identification}) problem --- without some shared prior framework
there is nothing that distinguishes Babe from an impersonator.
Specifically, other than ``eavesdropping'' Eve may pretend to be Babe
and trick Adam to tell her something that he would only tell Babe.
The use of a shared secret key for authentication reduces the quantum
cryptosystem to a key expansion scheme as noted in
Ref. \cite{bennet2}, without many advantages of a public-key system.
In particular, a separate key is needed for each pair of users which
causes major problems in a network environment.  In standard
cryptography there are a variety of approaches \cite{menezes} to
dispense with the use of shared secret keys, notably the use of
digital signature for identification that is capable of preventing the
identifier or verifier to pretend to be the identifiee.

In this paper a new cryptographic tool, {\em anonymous key encryption}
(AKE), is introduced in the quantum context that has no known parallel
in standard cryptography.  In AKE, the encrypter does not know the
value of his encrypted message.  If quantum states can be stored,
i.e., if quantum memory is available which is a subject of active
current effort, the AKE technique can be extended to a general
anonymous key technique that leads to various forms of digital
signature and to a public-key type identification protocol, to be
called {\em anonymous key identification} (AKI), which does not
require any shared secret key.  For key distribution, an unconditional
security proof on the use of AKE would be described for qubits, and it
would be indicated how a similarly secure protocol may be obtained in
the presence of noise and loss by using classical error correcting
codes.  The possible use of large-energy coherent states would also be
indicated.

Let $|\psi_A \rangle \in {\cal H}$, where $\cal H$ is an arbitrary
quantum state space, be a state known only to Adam and transmitted by
him to Babe.  Depending on the message $j\in \{1,\cdots, m\}\equiv
{\cal M} $ that Babe wants to send to Adam, she modulates $|\psi_A
\rangle$ with a unitary transformation $U^B_j$ and send $U^B_j |\psi_A
\rangle$ back to Adam.  From knowledge of $|\psi_A \rangle$ and the
openly known $U^B_j$, Adam can decrypt $j$.  The idea is that without
knowing $|\psi_A \rangle$, Eve cannot tell $j$ without significant
error.  The name anonymous key encryption is chosen because
$|\psi_A \rangle$ acts like an encryption key for Babe to generate the
encrypted signal $U^B_j |\psi_A \rangle$ with data $j$.  Often one has
$n$ qubits ${\cal H} = \bigotimes ^n{\cal H}_2$ with ${\cal M} =
\{0,1\}^n$.

Consider the following concrete AKE system for a single qubit $m = 2,
n = 1$, so that an arbitrary pure state $\rho_A = |\psi_A\rangle
\langle \psi_A |$ is represented in
terms of a real vector $\bar{r} \in {\bf R}^3$ via the Pauli
matrices  $\bar{\sigma}$, in component
form
\begin{equation}
\rho_A = \frac{1}{2}(I+r_1 \sigma_1+r_2 \sigma_2+r_3 \sigma_3),
\hspace{.3in}|\bar{r}|^2 =1
\end{equation}
\noindent
If $\rho_A$ is one of $M$ possible uniformly distributed
states on the $(\sigma_1, \sigma_3)$ great circle of the Bloch sphere
(or Poincare sphere in the context of photon polarization), we have
\begin{equation}
r_1 = \cos \frac{2 \pi \ell}{M}, r_2 = 0, r_3 = \sin
\frac{2 \pi \ell}{M} \hspace{.4in} \ell \in \{1,\ldots , M\}
\end{equation}
If $j=1$, Babe rotates $\rho_A$ by an angle $\frac{\pi}{2}$ clockwise
on this great circle, and if $j=0$, she rotates it by an angle
$\frac{\pi}{2}$ counterclockwise, i.e., $U^B_j = U(\phi_j)$, the
rotation matrix with $\phi_j = \pm \frac{\pi}{2}$.  These two states
are orthogonal in a basis known only to Adam, which he can measure to
determine $j$. Equivalently, the rotation angles may be $\{ 0, \pi \}$
or some other pairs.  In order that $U(\phi_j)|\psi_A \rangle$ is one
of the $M$ possible states $\rho_A (\ell)$ of $(1) - (2)$, $M$ is
taken to be a multiple of 4.  If Adam picks $\rho_A (\ell)$ randomly,
the resulting density operator $\rho_B = \sum_\ell \frac{1}{M}
U(\phi_j ) \rho_A (\ell )U^{\dagger} (\phi_j )$ from $B$ to $A$ is the
same for either $j$.  Thus, {\em even if} Eve has an identical
copy \cite{noclon} of the state sent back to $A$, she can gain no
information on $j$. This generalizes to a sequence of independent
$\rho_A^i (\ell)$ with independent $i$, for which Eve's optimal joint
attack on $\rho_B$ just factorizes into a product of individual
attacks.

The security analysis is carried out 
via the theory of optimal $M$-ary quantum
detector \cite{helstr,yuke} in which 1 out of $M$ possibilities, each
described by a state $\rho_j$ and a priori probability $p_j$, is
selected to optimize a given performance criterion.  The selection is
based on the result of a general quantum measurement described by a
positive operator-valued measure (POM), which is specified to yield
the optimal performance.  If Eve attempts to identify $\rho_A (\ell)$
by intercepting the transmission to Babe, the best she can do is given
by the optimum $M$-ary quantum detector for discriminating the states
$(1) - (2)$, which has been worked out before.  Lemma 3 of Ref.
\cite{yuke} gives the optimum quantum measurement 
in the form of a POM, with
corresponding probability of correct identification given by
$P_c^{\prime} = 2/M$.  However, even if Eve makes an error, her
estimated state is still useful for eavesdropping purpose and a
different criterion needs to be used.  Generally, it is the
probability $P_a$ that Eve's estimated state is accepted to be correct
by Adam as a result of his measurement.
\begin{equation}
P_a = \sum_{\ell, \ell^{\prime}} \frac{1}{M} p(\ell^{\prime} |\ell)
{\rm tr} \rho_A
(\ell) \rho_A^E (\ell^{\prime})
\end{equation}
where $p(\ell^{\prime} | \ell ) ={\rm tr} \Pi (\ell^{\prime}) \rho_A(\ell
)$ is the probability that given $\rho_A(\ell)$ was transmitted, Eve
takes it to be $\rho_A^E (\ell^{\prime} )$ from measuring the POM
$\Pi(\ell^{\prime} )$.  Such a criterion falls under the general
optimum quantum detector formulation, and the optimum
$\Pi(\ell^{\prime})$ for (3) turns out \cite{note1} to be the same as that
of determining $\rho_A (\ell)$ according to the error probability
criterion, which is intuitively reasonable.  The resulting $P_a$ is
given by 3/4 independently of $M$ (but recall that $M$ is a multiple
of 4). Thus, if Eve measures $\Pi (\ell^{\prime})$ to determine
$\rho_A (\ell)$, perhaps because she cannot store the actual $\rho_A
(\ell)$, and transmits $\rho_A^E (\ell^{\prime})$ to Babe, determines
$j$ by measurement on $U(\phi_j ) \rho_A^E (\ell^{\prime} )
U^{\dagger} (\phi_j)$ from Babe, and sends the resulting state back to
Adam, the probability that Adam decrypts correctly is 3/4.  Pure
guessing without measurement yields $P_a = 1/2$.  This $P_a = 3/4$
would be reduced to 2/3 if the whole Bloch sphere is utilized, with
$\rho_A$ given by (1) with $r_1 = \sin \theta \cos \phi , r_2 = \cos
\theta, r_3 = \sin\theta \sin\phi$, and e.g., $U^B_j = U(\theta_j),
\theta_j = \pm \frac{\pi}{2}$ with $\phi$ unchanged.  In the case (2),
$M = 4$ is enough to yield $P_a = 3/4$, and in this case a total of $M
= 6$ states \cite{brub} on the poles of any rectangular coordinate system
intercepting the Bloch surface would yield $P_a = 2/3$. In both cases
Eve can get these values of $P_a$ without knowing $M$ by measuring an
orthogonal basis chosen randomly from the $M$ possible
states. Evidently $P_a$ can be further reduced if a higher dimensional
$\cal H$ is used. 

If Eve could intercept and store $\rho_A (\ell)$, she could eavesdrop
perfectly by sending her own $\rho_E (\ell ')$ to Babe in an
impersonation attack.  Such manipulation can be detected with test
qubits mixed into the information qubits.  However, a different
approach is employed here in which Babe sends her modulated qubits
back to Adam in a random order.  This has the advantage that all
possible eavesdroppings can be thwarted without checking for
disturbance, thus allowing a simple proof of protocol security for key
establishment.  In this scheme, Adam and Babe use AKE with $8k$ qubits
to establish a key of length $4k$ while expending a shared secret key
of length $2k$, resulting in a net key expansion of $2k$ as follows.
For each 8 qubit block, Babe sends back the qubits in one of the
following four orders equiprobably using 2 secret bits: 12345678,
87654321, 38462715, 41236587.  These four sequences are chosen so that
there is no qubit overlap in any position among the eight.  Eve can
alter the qubits from A to B in an impersonation attack, or to conduct
opaque eavesdropping, or to conduct translucent eavesdropping by
tapping into the communications between A and B to learn about $j$.  The
probability that Eve guesses $q$ of the $k$ qubit groups in the right
order in an impersonation attack is given by the binomial distribution
with success probability 1/4, and thus is exponentially small in $q$.
The rest she induces an error probability $P_e = 1/2$  per qubit
for Adam, and the key establishment would fail in a trial encryption.

Eve may employ an opaque eavesdropping strategy by intercepting and
re-transmitting the states from A to B and B to A.  Instead of using
disturbance detection, we merely use classical privacy amplification
(CPA) \cite{bbcm} to eliminate Eve's partial information.  Eve's
success probability $P_c$ per qubit is bounded as follows.  We grant
her one copy of $\rho_A (\ell)$ and one copy of the corresponding
correct $U (\phi_j) \rho_A (\ell ) U^\dagger (\phi_j )$, i.e.,
we allow Eve to intercept both copies exactly as if there is no
disturbance and the order is correct.  From these two copies she can
try to learn $j$ by optimally processing both states.  This is a
binary detection problem with two states

\begin{equation}
\rho_{0,1} = \sum^M_{\ell = 1} \frac{1}{M} \rho_A (\ell ) \rho_A \left(\ell
\mp \frac{M}{4}\right)
\end{equation}
for which the optimum probability of discrimination can be obtained by
diagonalizing $\rho_0 - \rho_1$ \cite{helstr}.  The resulting optimum
probability $P_c$ she would determine $j$ correctly turns out to be
the same as that obtained by measuring the optimum state detector on
the copy $\rho_A (\ell)$ from A to B and then measuring whether
the state $\rho_B (\ell)$ from B to A is clockwise or
counterclockwise with respect to $\rho_A (\ell)$, which is intuitively
reasonable, and is given by $ P_c = P_a$.  If Eve launches a joint
attack by making measurements on blocks of qubits, she cannot obtain a
better accuracy than that of measuring one by one --- the optimum
quantum detector for the bit error sum factorizes when both the states
and the data probabilities of the blocks factorize into a product from
the corresponding bits.

In translucent eavesdropping, Eve would try to determine the data $j$ by
correlating her tappings from A to B and B to A.  She can do this in
the correct order only with probability 1/4.  Thus, to (loosely) bound
all the possible information Eve can obtain, we let her succeed in
learning the bits exactly with probability 1/4, and with probability
3/4 we let her learn the bits with probability $P_c = P_a$ as in (4)
above.  For $P_a$ = 3/4, this yields a total of $2k$ deterministic bits
and $<k$ Shannon bits, which can be eliminated by expending $4k$ bits
or just $3k$ bits asymptotically \cite{bennet2, bbcm}.  This completes
the security proof in the ideal limit.

Note that no quantum memory is required in this scheme.  We have used
very loose bounds to avoid complex arguments and bounding techniques,
but the resulting efficiency is still appreciable.  The present AKE
has no apparent classical analog because listening to both the
transmissions from A to B and B to A would reveal too much about the
bits in a classical system even when the bit order is random.  The
intrinsic statistical feature of quantum ontology, that it is
impossible to determine the state of a single quantum system exactly,
is directly expoited in AKE.  The basic ingredients of our security
guarantee are: use of qubit order randomization to thwart manipulation
and correlation, use of optimum quantum detector and copies to Eve to
bound her partial information which is eliminated by classical privacy
amplification, and use of classical error correcting code to overcome
loss and noise to be presently discussed.  In particular, the explicit
use of a shared secret key for key expansion, in this case in
obtaining secret qubit orders, is a {\em new} technique that I expect
to be widely applicable in many scenarios.

The major problem for quantum security proof lies in the presence of
loss and noise in realistic systems.  It should be clear that the
above security proof does not depend on detecting small disturbance by
Eve, and can thus be expected to work in a similar way in the presence
of small noise and loss with some simple error correction capability.
In particular, one may employ classical error correcting codes (CECC)
on qubits in lieu of quantum codes.  Thus, each codeword in a CECC
$(x_i )$, $x_i \in \{0,1\}$, becomes a codeword of quantum states
$(|x_i \rangle)$, where $|x_i \rangle$ is the state corresponding to
$0$ and $1$ in the quantum modulation scheme adopted.  No
reconciliation\cite{bennet2} is needed with the use of CECC.

The protocol  for  AKE key distribution is in general:
\begin{itemize}

\item[(i)] Adam sends enough randomly chosen $|\psi_A \rangle$'s to
Babe to cover the loss and noise in transmission to Babe as well as
the CECC Babe needs to use for transmission back to Adam.

\item[(ii)]Babe modulates the information qubits in a known CECC, sends
the resulting qubits to Adam in a random order according to a short
shared secret key.

\item[(iii)]Some form of CPA is employed by
Adam to eliminate
any possible leakage of information which is strictly bounded.

\item[(iv)] The resulting key is checked for correctness by a trial
encryption. 
\end{itemize}

\noindent There are many variations of this protocol including the use of test
qubits or quantum memory in lieu of shared secret key.  There are also
many ways to use AKE for direct encryption.  These topics and the
security proof of the above protocol will be developed elsewhere.  

If quantum states can be stored, some features of public-key
cryptography can be obtained as follows.  In classical public-key
cryptography, a one-way function $f$: X $\rightarrow$ Y is roughly a
map for which one can obtain $y = fx \in$ Y from $x \in$ X readily but
it is ``infeasible'' to obtain $x$ from $fx$.  A one-way trapdoor
function results if $x$ can be readily obtained from $fx$ with
additional ``trapdoor information'' relating to $f$ \cite{note3}.  For
a physically given $|\psi_A \rangle$, the function ${\cal M}
\rightarrow {\cal H}$ with $j$ mapped into $U^B_j|\psi _A \rangle$ can
be regarded as a {\em quantum one-way function} with trapdoor
information given by the knowledge of the actual state $|\psi
_A\rangle$, to be denoted $K\psi _A$.  Thus, $|\psi _A\rangle$
functions like a {\em quantum public key} while $K\psi _A$ is the
private key.  Similar to the usual one-way trapdoor function, one can
obtain the physical state $U^B_j|\psi _A \rangle$ with a given public
key $|\psi _A\rangle$, but cannot obtain from $U^B_j|\psi _A \rangle$
the value $j$ without the knowledge $K \psi _A$.  This is the general
formulation of the anonymous quantum key technique.  It is clear that
AKE can be described in this way, with Adam sending Babe his public
key $|\psi _A\rangle$ and Babe using $|\psi _A\rangle$ to encrypt a
message $j$ which only Adam can decrypt with $K\psi _A$.  With $|\psi
_A\rangle$ representing a sequence of qubits, a number of standard
public key protocols can be recasted in the quantum domain.  For
example, one-time digital signatures and blind
signatures \cite{menezes} can be implemented this way.  Here, we would
use the anonymous key technique to obtain a quantum identification
protocol AKI of the challenge-response type in which the identifier
cannot pretend to be the identifiee and which is an exact analog of a
protocol \cite{menezes2} based on classical digital signature.  In AKI
Adam uses his stored $|\phi_B \rangle$, $\phi_B$ unknown to him, to
identify Babe in the following way.  He modulates $|\phi_B \rangle\in
{\cal H}_2$ with a randomly chosen $\phi_A$ and transmits $|\phi_B +
\phi_A \rangle$ to Babe, say for states of the form (1) - (2), and
asks her to return the state $|\phi_A \rangle$with $\phi_B$ removed
which Babe is capable of doing by just adding $ - \phi_B$ to the angle
in $|\phi_B + \phi_A \rangle$. Adam checks by measuring the projection
to $| \phi_A \rangle$.  The random $\phi_A$ is necessary, or else Eve
can just return the state $|\phi = 0 \rangle$, where $\phi = 0$ is the
reference angle, without using $|\phi_B \rangle$ sent by Adam. The
protocol can be simply summarized:
%%
%% wlk
%%
\begin{equation}
\begin{array}{llll}
  &  \mbox{(i)} & A \rightarrow B: & | \phi_B + \phi_A \rangle  \nonumber \\
  &             &                  &  \\
  &  \mbox{(ii)} & B \rightarrow A: & | \phi_A \nonumber \rangle
\end{array}
\end{equation}
The probability that Adam or Eve could successfully impersonate Babe
is $P_a$ for one qubit, which can be brought to any desired security
level $P_s = (P_a)^m$ with $m$ qubits exponentially efficiently.
Apart from using quantum laws instead of number theoretic complexity
assumptions, the security of this protocol is evidently the same as
the conventional public-key challenge-response indentification
protocol \cite{menezes2}.  Note that the success of AKI is independent
of that of AKE, with both being examples of the anonymous quantum key
technique.  

This technique can also be used to obtain unconditionally secure
quantum bit commitment schemes, outside the framework of the
impossibility proof \cite{lo}, which is not sufficiently general to
rule out all such schemes.  In one of these, Babe sends anonymous
states to Adam for bit modulation and the anonymous nature of the
states prevents Adam from determining the cheating unitary
transformation on his committed state.  In another, the anonymous
states prevent both Adam and Babe from cheating.  A detailed treatment
of quantum bit commitment is given in ref \cite{yuenun}.

Some comments on possible experimental realization are in order.  If
(1)-(2) are realized via photon number polarization, a small $M$ is
sufficient as indicated after Eq.~(3).  Although our protocol is much
simpler, the experimental setup would be quite similar to BB84, and
the efficiency would suffer greatly in the presence of loss.  In the
present M-ary approach, however, it can be improved via large-energy
coherent states by the use of a further new technique to be elaborated
elsewhere.  One
underlying reason for such possibility can be explained. Consider the
coherent states

\begin{equation}
|\alpha_0(\cos \theta_{\ell} + i \sin \theta_{\ell} \rangle ,
 \hspace*{.3in} \theta_{\ell} = \frac{2 \pi \ell}{M}
\end{equation}
for a real positive $\alpha_0$ in place of $(1)-(2)$.  Any two basis
states of (6) have inner product $exp (-2\alpha_0^2 )\sim 0$ for large $\alpha_0$.  When $M \rightarrow \infty$ or when $M$ is unknown,
one obtains $P_a = 1/2$ with heterodyne detection and $P_a < 2/3$ for
the canonical phase measurement which is the maximum likelihood phase
estimator \cite{holevo}.  This important behavior of having $P_a$
independent of $\alpha_0$ would also be obtained for a known finite $M
\gg \alpha_0$, as a lower bound to the mean-square fluctuation
$(\delta \theta)^2$ was obtained \cite{yuen3} that goes as $1/|\alpha
|^2$ for coherent states $|\alpha \rangle$.  In the $P_a$ expression,
this fluctuation would cancel out the $\alpha_0 ^2$ in the form $2
\alpha_0^2 \sin^2 \frac{\delta\theta}{2}$ when $M \geq 2\pi /
\delta\theta$.  A two-mode coherent state realization similar to (6),
with $|\alpha_0 \cos \theta_{\ell} \rangle |\alpha_0 \sin
\theta_{\ell}\rangle$, can also be used.  In either case $M \sim 10^3$
is easily achievable in the laboratory, with much higher $M \geq 10^6$
possible, so that large $\alpha_0$ can be used for overcoming loss and
noise.  This is not possible in previous quantum cryptosystems such as
modified BB84 because large $\alpha_0$ would lead to unambiguous
determination of the states involved, which is not the case if there
are many states $M \gg \alpha_0$. The use of (6) also allows the
possibility of amplification and regeneration along the transmission
path using quantum amplifiers \cite{yuen2000}, as well as routing and
switching in a network.  Analysis of such coherent-state systems
will be given in a future publication detailing how key distribution
and encryption can be carried out.  It appears that they hold great
promise in making secure quantum cryptography truly practical.


\begin{thebibliography}{99}
\bibitem{bb84}  For the BB84 protocol, a public
non-jammable channel would prevent Eve from eavesdropping  via
impersonation  if Adam and Babe can identify themselves.
\bibitem{note1} The cost $P_a$ to be optimized can be put into the same
form as $P_c^{\prime}$, but with a new $\rho_A^{\prime}(\ell)$ that is
related to $\rho_A (\ell)$ of $(1)-(2)$ by a constant factor of 2 on
$\bar{r}$, which does not affect the optimal detector solution.
\bibitem{note3} The famous RSA encryption function involves $fx = x^b$ for
a positive integer $b$, X = Y = set of integers modulo $n$, $n = pq$
the product of two large primes.  An $f$ properly chosen this way is
believed to be one-way, due to the computational complexity of
factoring $n$ with $p,q$ being the trapdoor information.
\end{thebibliography}
\end{document}